\documentclass[a4paper]{article}
\usepackage{graphicx}
\usepackage[style=numeric-comp,useprefix,hyperref,backend=bibtex]{biblatex}
\bibstyle{plain}
\bibliography{bibelectromechanics}
\usepackage{graphicx,float}
\usepackage{amsmath}
\usepackage{esint}
\usepackage{color}
\usepackage{MACROS-q}
\renewcommand{\bel}[1]{  \begin{equation} \label{#1}}
\renewcommand{\eel}{\end{equation}}
\renewcommand{\derp}[2]{\displaystyle  \frac{\partial\;\! #1}{\partial\;\! #2}}
\renewcommand{\fra}[2]{\displaystyle \frac{#1}{#2}}

\renewcommand{\der}[2]{\displaystyle \frac{\mbox{d}\;\!#1}{\mbox{d}\;\!#2}} 
\renewcommand{\dd}{\mbox{d}}
\usepackage{amssymb}
\usepackage{mathbbol,bbm}

\usepackage{bm,dsfont}

 \def\F{{\mathbb{F}}}

\def\dd{\mathord{\rm d}}

\usepackage{booktabs}

\usepackage{xcolor}
\usepackage{amsfonts}
\usepackage{tikz}
\usetikzlibrary{calc}
\usetikzlibrary{trees}
\usetikzlibrary{snakes}
\usetikzlibrary{automata}
\usetikzlibrary{arrows}
\usetikzlibrary{positioning}

\begin{document}
\title{Electromechanical power flowcharts in systems of electrical circuits}
\author{J.M. Diazdelacruz \\Department of Applied Physics and Materials Engineering\\ Universidad Politecnica de Madrid, Spain  }

\maketitle
\tableofcontents

\begin{abstract}
We present an original undergraduate level compilation for the physics of electromechanical systems with special attention to power flow. An approach based on energy considerations is presented  that is specially suited to compute the mechanical and electrical actions of electromagnetic fields and to draw power flowcharts that clarify the path taken by energy in typical devices. The procedure guarantees energy conservation and provides a consistent way for auditing the power flow.
\end{abstract}




\section{Introduction}\label{sec::intro}
Classical textbooks on electromagnetics \cite{corson},\cite{griffiths},\cite{hayt},\cite{reitzmilford},\cite{sadiku} currently taught in undergraduate engineering courses often place great emphasis on electromagnetic induction. It is not surprising, since it plays a paramount role in energy transport and electromechanical conversion in today's technology. Yet, in our perspective, a more thorough analysis of the interplay of  electrical, mechanical and magnetic variables in a typical machine would also fit into an undergraduate engineering course on electromagnetics. It is our intention to sketch the theory of such a systems placing a special focus on power flow tracking and charting. On the other hand, texts on electromechanical systems, often appeared in a Mechatronics context\cite{Crandall, Preumont} make use of analytical mechanics to provide the mathematical framework for a generalized formulation of the subject. We propose a way in between the raw Lorentz force and the analytical mechanics approach, which is better suited for students with no previous course on theoretical mechanics. It is bases on power balances which, in turn, rely on the conservative character of the energy in electromagnetic systems in the low frequency regime, under which most electromechanical systems operate.

We consider electromagnetic currents at low frequencies, meaning that signals propagating at the speed of light do not experience a significant change of phase when they come through the electric machine. We will further assume that conduction currents  flow in closed paths (circuits). They may move, typically rotate about a fixed axis, and accordingly, dynamical actions generate a mechanical power flow. Faraday's law, on its turn, induces electromotive forces in circuits, so that there is also electrical power.  

This lesson aims at providing a systematic way for computing mechanical and electromotive forces and power flows and check that they are consistent with energy conservation. We do it both analytically and graphically.

\section{Electromagnetic potential energy}\label{sec::empe}
Considering the volume density of power delivered by the electromagnetic field to the conduction currents
\bel{prim}
p=\vec J\cdot\vec E
\eel
it is possible to write an overall power balance
\bel{oallpb}
\iiint p \dd \tau = - \iiint \vec H \cdot \derp{\vec B}t \dd \tau - \iiint \vec E \cdot \derp{\vec D}{t} \dd \tau - \oiint (\vec E\times\vec H) \,\cdot \dd \vec \sigma  
\eel
where, assuming constitutive relations $\vec D=\vec D(\vec E)\,,\,\vec B=\vec B(\vec H)$ \footnote{it is further assumed that $\derp{B_i}{H_j}=\derp{B_j}{H_i}\,,\,\derp{D_i}{E_j}=\derp{D_j}{E_i}$}, and defining\footnote{Vector $\vec S$ is known as {\em Poynting Vector}}
\bel{defis}
\lliz{rcl}
u_e &:=&\int_{\vec{0}}^{\vec D} \vec E\cdot\dd \vec D\\
\\
u_m &:=&\int_{\vec{0}}^{\vec B} \vec H\cdot\dd \vec B\\
\\
\vec S&:=&\vec E\times\vec H
\clliz
\eel
it follows that
\bel{lpb}
\iiint p \dd \tau = - \iiint \derp{(u_e+u_m)}{t} \dd \tau - \oiint \vec S \,\cdot \dd \vec \sigma  
\eel
that renders physical interpretations of $u_e,u_m,\vec S$ as electric and magnetic energy densities and current respectively.

Slowly varying electric charges and currents yield electric and magnetic fields that decay at most as $1/r^2$, so that the flux of the Poynting vector $\vec S$ becomes negligible if the integration is taken over a big sphere. That allows us to define a potential energy $U_{\rm em}$ from which conduction currents take or deliver power.

\bel{empe}
U_{\rm em}:=\iiint_{\rm all\,space} (u_e+u_m)\dd \tau
\eel

In linear and isotropic materials $\vec D=\epsilon\vec E, \vec B=\mu \vec H$, where $\epsilon,\mu$ do not depend on the fields. Then, 

\bel{linears}
\lliz{rcl}
u_e &=&\frac 1 2 \epsilon |\vec E|^2 \\
\\
u_m &=&\frac 1 {2\mu}  |\vec B|^2\\
\\
U_{\rm em}&=&\frac 1 2 \iiint_{\rm all\,space} (\epsilon |\vec E|^2+\frac{1}{\mu} |\vec B|^2)\dd \tau
\clliz
\eel

\section{Electromotive and mechanical forces on circuits}\label{sec::emf}

Next, we set the scenario of electromechanics with currents. A set of $n$ electric circuits, each one being traversed by a magnetic flux $\Phi_j\,j=1,\ldots,n$, is positioned by a collection of $m$ generalized geometrical coordinates $q_1,\ldots,q_m$. Each electric circuit is given energy at a rate  $- I_j \dot \Phi_j$ and every mechanical degree of freedom receives a power $F_k\dot q_k$, where $F_k$ is the generalized force corresponding to the $k-$th generalized coordinate $q_k$. As the action of the electromagnetic field on conduction currents is conservative, it follows that
\bel{bal}
\der{U_{\rm em}}{t}=\sum_{j=1}^n I_j \dot \Phi_j-\sum_{k=1}^m F_k \dot q_k  
\eel
whence, if energy $U$ is expressed as a function of the fluxes and coordinates $U=F(\Phi_j,q_k)$, we readily arrive at
\bel{r}
\lliz{rcl}
I_j&=&\derp F {\Phi_j}\\
\\
F_k&=& -\derp F {q_k}
\clliz
\eel 
Eq.\ref{bal} implies a way to compute $U_{\rm em}$ by raising the set of currents from zero to their final values keeping the positions constant
\bel{comp}
U(I,q)=\sum_{j=1}^n \int_0^\Phi I_j \dd \Phi_j(I,q) = \sum_{j=1}^n I_j\Phi_j(I,q)- \sum_{j=1}^n \int_0^I \Phi_j(I,q) \dd I_j
\eel
where the sets of $I_1,\ldots,I_n$ currents and $q_1,\ldots,q_m$ coordinates are denoted generically as $I,q$. It is often useful to perform a Legendre transformation and define the {\em co-energy} $W$ as
\bel{coenergy}
W:=\sum_{j=1}^n I_j\Phi_j - U=\sum_{j=1}^n \int_0^I \Phi_j(I,q) \dd I_j
\eel
and express it as a function of the currents and the coordinates $W=G(I_i,q_k)$, so that
\bel{ra}
\lliz{rcl}
\Phi_j&=&\derp G {I_j}\\
\\
F_k&=& \derp G {q_k}
\clliz
\eel 
When $F(\Phi_j,q_k)$ is a quadratic function of the fluxes, energy and co-energy are equal  $U=W$. This is the case in linear systems.

The electromotive force on circuit $j$ is always given by Faraday's law
\bel{faraday}
{\cal E}_j = - \der{\Phi_j}{t}
\eel

\section{Generalized forces on circuits from external magnetic fluxes}\label{sec::ext}

In this section we consider the mechanical actions on a set $C_1$ of $n_1$ electric circuits originated by fields created by other set $C_2$ of  $n_2$ currents which are not exactly known or we are not interested in. It is assumed that the magnetic flux through the $\ell$-th circuit in $C_1$ can be split into two contributions: $\Gamma_j, \Psi_j$, from $C_1,C_2$, respectively.  All we need is an expression for the fluxes $\Psi_{\ell j}(q_k,t)\,;\,\ell=1,\ldots,n\,;\,j=1,\ldots,n$ through the $\ell$-th circuit in $C_1$ caused by the $j$-th current in $C_2$.  From Eq.\ref{r},
\bel{re}
F_k = - \derp{F}{q_k}
\eel
where
\bel{f}
F(\Phi_i,q_k)=\sum_{j\in C_1\cup C_2} \int_{0}^{\Phi} I_j(\phi,q_k) \dd \phi_j=\sum_{j\in C_1\cup C_2}\Phi_j I_j(\Phi,q_k)-\sum_{j\in C_1\cup C_2}\int_{0}^{I(\Phi,q_k)} \Phi_j(i,q_k) \dd i_j
\eel
whose partial derivative yields
\bel{reh}
F_k = - \derp{F}{q_k}=\sum_{j\in C_1\cup C_2}\int_{0}^{I(\Phi,q_k)} \derp{\Phi_j(i,q_k)}{q_k} \dd i_j
\eel
If we consider mechanical actions on a coordinate $q_k$ that positions circuits in $C_1$ from fields created by circuits in $C_2$, the currents taken into account are only those in $C_1$ and the fluxes $\Psi_j$ from $C_2$ do not depend on said currents. Thus,
\bel{reh}
F_k = \sum_{\ell=1}^{n_1}\int_{0}^{I_\ell(\Phi,q_k)} \derp{\Psi_{\ell}(q_k,t)}{q_k} \dd i_\ell= \sum_{\ell=1}^{n_1}{I_\ell} \derp{\Psi_{\ell}(q_k,t)}{q_k} 
\eel
and the mechanical power delivered to circuits in $C_1$ is
\bel{rehp}
P_{m,C_1} = \sum_{k=1}^m  \sum_{\ell=1}^{n_1}{I_\ell} \derp{\Phi_{\ell}(q_k,t)}{q_k}  \dot q_k = \sum_{k=1}^m  \sum_{\ell=1}^{n_1}{I_\ell} \derp{[\Psi_{\ell}(q_k,t)+\Gamma_{\ell}(q_k,t)]}{q_k}  \dot q_k
\eel
whereas the electrical power evaluates to
\bel{rehpe}
P_{e,C_1} = \sum_{k=1}^m  \sum_{\ell=1}^{n_1}{I_\ell} \der{\Phi_{\ell}(q_k,t)}{t} = \sum_{k=1}^m  \sum_{\ell=1}^{n_1}{I_\ell} \der{[\Psi_{\ell}(q_k,t)+\Gamma_{\ell}(q_k,t)]}{t} 
\eel
When all $\Psi_j(q_k,t)$ do not depend on time, their contribution to the sum of Eqs.\ref{rehp},\ref{rehpe} cancels, meaning that they do not contribute any power to $C_1$, and they act just as {\em catalyst} for electro-mechanical energy conversion.

It is possible to define a {\em partial} potential energy for the circuits in $C_1$.
\bel{loc}
U_p:= \sum_{j=1}^{n_1} \int I_j \dd \Gamma_{j}=\sum_{j=1}^{n_1} I_j \Gamma_j(I,q) -\sum_{j=1}^{n_1} \int_0^I \Gamma_j(q,i)\dd i_j
\eel
whose time derivative is
\bel{loce}
\der{U_p}{t}= \sum_{j=1}^{n_1} \der{I_j}{t} \Gamma_j(I,q) +\sum_{j=1}^{n_1} {I_j}\der{ \Gamma_j(I,q)}{t} -\sum_{j=1}^{n_1}  \Gamma_j(q,I)\der{I_j}{t} -\sum_{j=1,k=1}^{n_1,m}  \int_0^I \derp{\Gamma_j(q,i)}{q_k}\dd i_j\,\dot q_k
\eel
which, combined with Eqs.\ref{rehp},\ref{rehpe} yields
\bel{qloc}
\der{U_p}{t}+P_{m,C_1}+P_{e,C_1}=0
\eel
so that, when external fluxes are constant, electrical and mechanical powers come only at the expense of the partial potential $U_p$.

Finally, as a particularization of Eq.\ref{reh}, it is worth noting that when $q_k$ determines the  position the $j$-th circuit the mechanical action of an external field whose magnetic flux through the circuit is $\Psi(q_k)$ reads
 \bel{rehf}
F_k = {I_j} \derp{\Psi_j(q_k,t)}{q_k} 
\eel

\section{Power flowchart}\label{sec::audit}
In this section we write down expressions for energy flows in electromechanical systems. Energy is conserved and accordingly positive and negative fluxes should balance out. The rate of decrease of the potential energy $U_{\rm em}$  should match the power delivered by the electromagnetic field to the electric circuits and the mechanical degrees of freedom.  We next check that this is the case. The $j$-th electric circuit receives a power from the electromagnetic field $P_{e,j}$ given by
\bel{pfar}
P_{\rm e,j}=I_j {\cal E}_j = -I_j\der{\Phi_j}{t} 
\eel
The power supplied to the $k-$th mechanical degree of freedom $P_{\rm m,k}$ is
\bel{pfadr}
P_{\rm m,k}=F_k\dot q_k  = \derp G {q_k}\dot q_k=\sum_{j=1}^{n} I_j\der{\Phi_j}{t} -\sum_{j=1,k=1}^{n,m}
\int_0^I \derp{\Phi_j(I,q)}{q_k}\dd i_j \dot q_k
\eel
Deriving Eq.\ref{comp}, one obtains
\bel{rdsa}
\der{U}{t} = \sum_{j=1}^{n} \der{I_j}{t}\der{\Phi_j}{t} +\sum_{j=1}^{n} I_j\der{\Phi_j}{t} -\sum_{j=1}^{n} I_j\der{\Phi_j}{t} -\sum_{j=1,k=1}^{n,m}
\int_0^I \derp{\Phi_j(I,q)}{q_k}\dd i_j \dot q_k
\eel
which, combined with Eqs.\ref{pfar},\ref{pfadr}, yields
\bel{nb}
\der{U_{\rm em}}{t} + \sum_{j=1}^n P_{\rm e,j} + \sum_{k=1}^m P_{\rm m,k} = 0
\eel
that is another statement of Eq.\ref{bal}. 

\begin{figure}[h]
\begin{center}\includegraphics[width=0.4\textwidth]{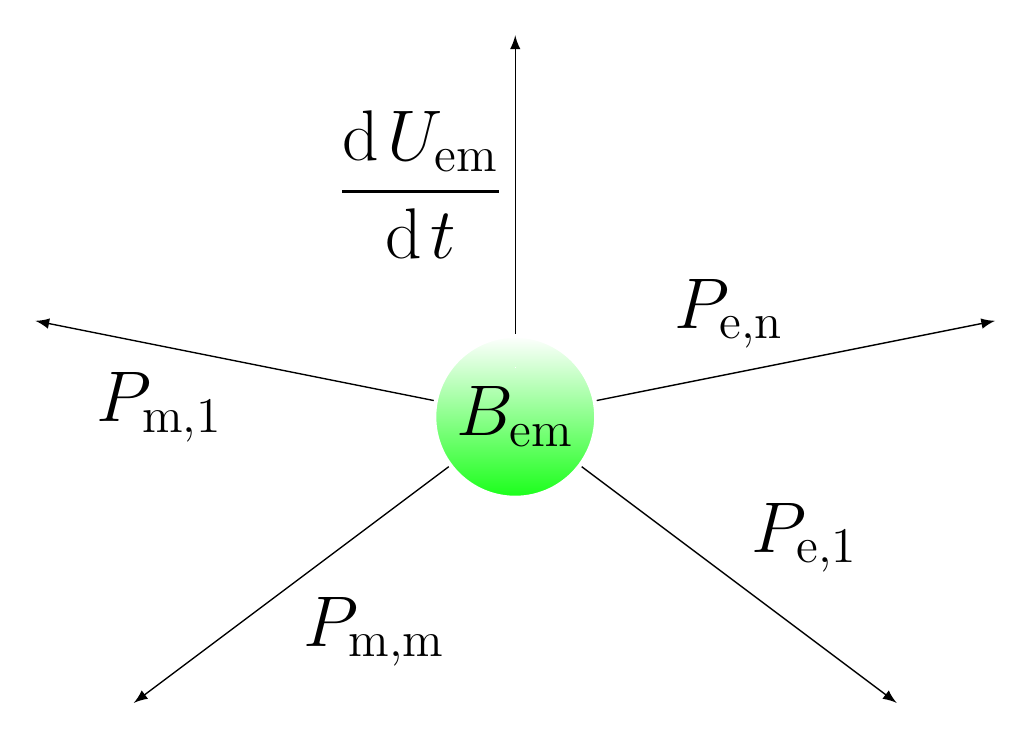}\end{center}
\caption{Graph representation of the power balance equation $\der{U_{\rm em}}{t} + \sum_{j=1}^2 P_{\rm e,j} + \sum_{k=1}^2 P_{\rm m,k} = 0$ that provides a visual picture of energy conservation.}
\label{fig::node}
\end{figure}

An equation in which an addition of powers equals zero is a {\em power balance equation}. It can be drawn as node in a graph if the edges represent the powers.  Fig.\ref{fig::node} depicts the power balance  Eq.\ref{nb} in the case $n=2, m=2$. 

\begin{figure}[h]
\begin{center}\includegraphics[width=0.75\textwidth]{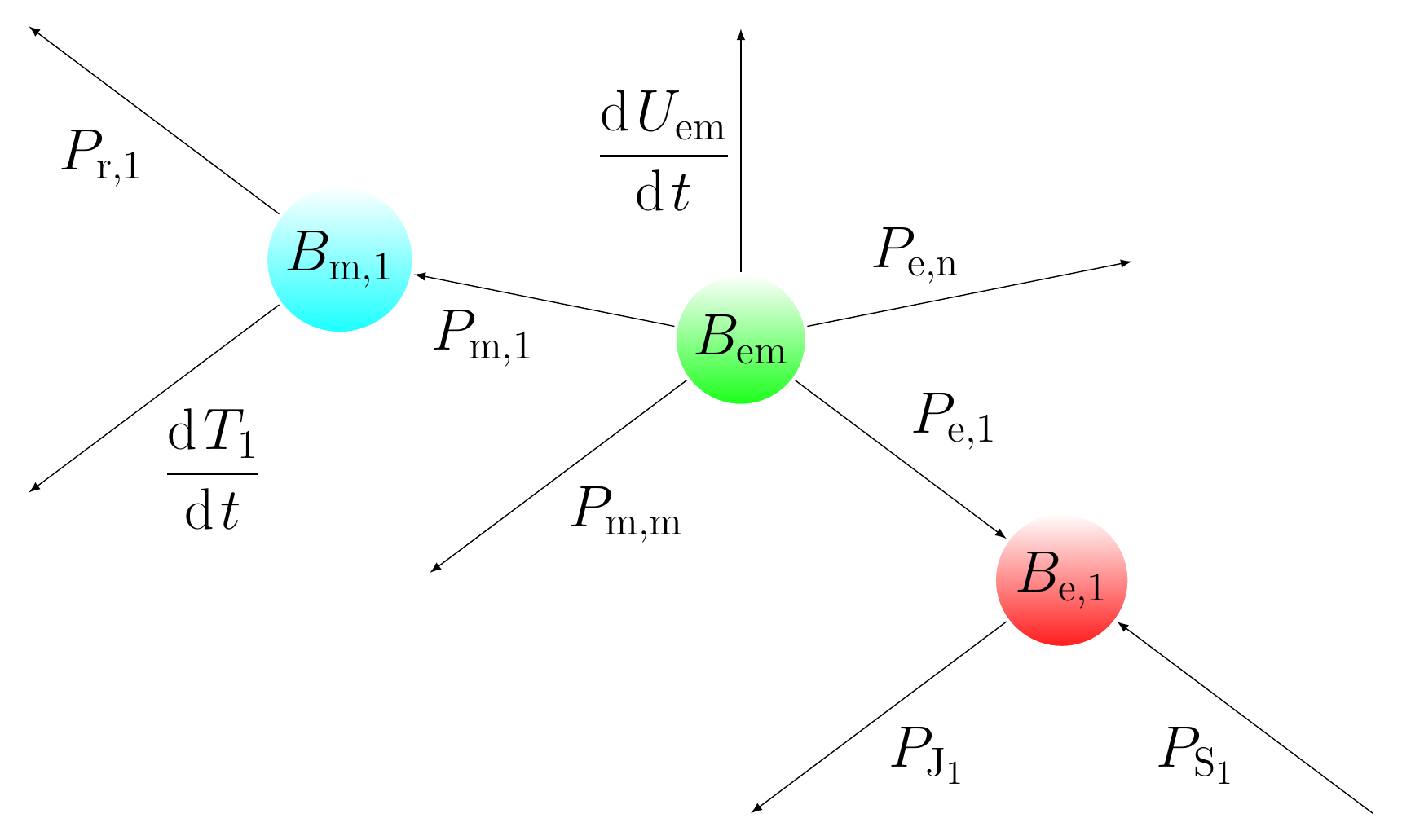}\end{center}
\caption{Flowchart of the path followed by the electrical and mechanical powers power $P_{\rm e,1},P_{\rm m,1}$ comming from Eq.\ref{nb}. $P_{\rm e,1}$ combines with power from an electrical source  $P_{\rm S,1}$ to be dissipated  as Joule power $P_{\rm J,1}$ by a resistor, whereas the mechanical power is used to overcome a mechanical friction $P_{\rm r,1}$ and increase a kinetic energy $T$.}
\label{fig::elmec}
\end{figure}

All  mechanical and electrical powers can be traced further to the mechanical or electrical systems  they are fed to or drawn from. Fig.\ref{fig::elmec} amplifies the scope of the flowchart to include information about the next stages of the power. For example, the electrical power delivered by the field to the first circuit $P_{\rm e,1}$ enters a new balance in which it is added to the power coming from an electric source $P_{\rm S,1}$ to feed a resistor where $P_{\rm J,1}$ is dissipated. The mechanical power that flows into the first geometrical degree of freedom also enters a mechanical power balance that splits it into a power dissipated by mechanical friction $P_{\rm r,1}$ and the time derivative of the kinetic energy or a moving part.

\begin{figure}[h]
\begin{center}\includegraphics[width=0.75\textwidth]{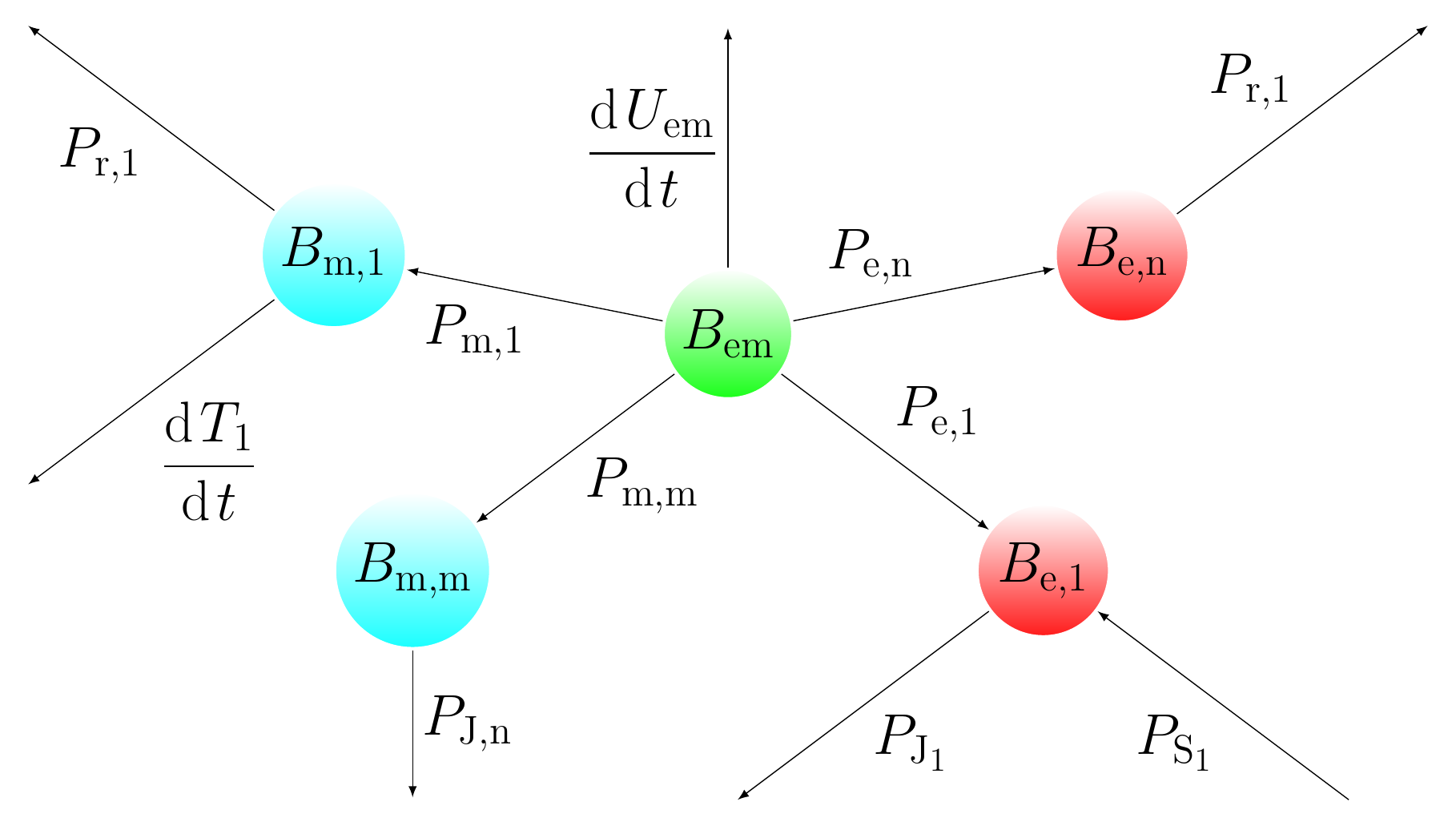}\end{center}
\caption{Possible complete flowchart of power in a $n=2,m=2$ electromechanical system.}
\label{fig::all}
\end{figure}

\begin{figure}[h]
\begin{center}\includegraphics[width=0.75\textwidth]{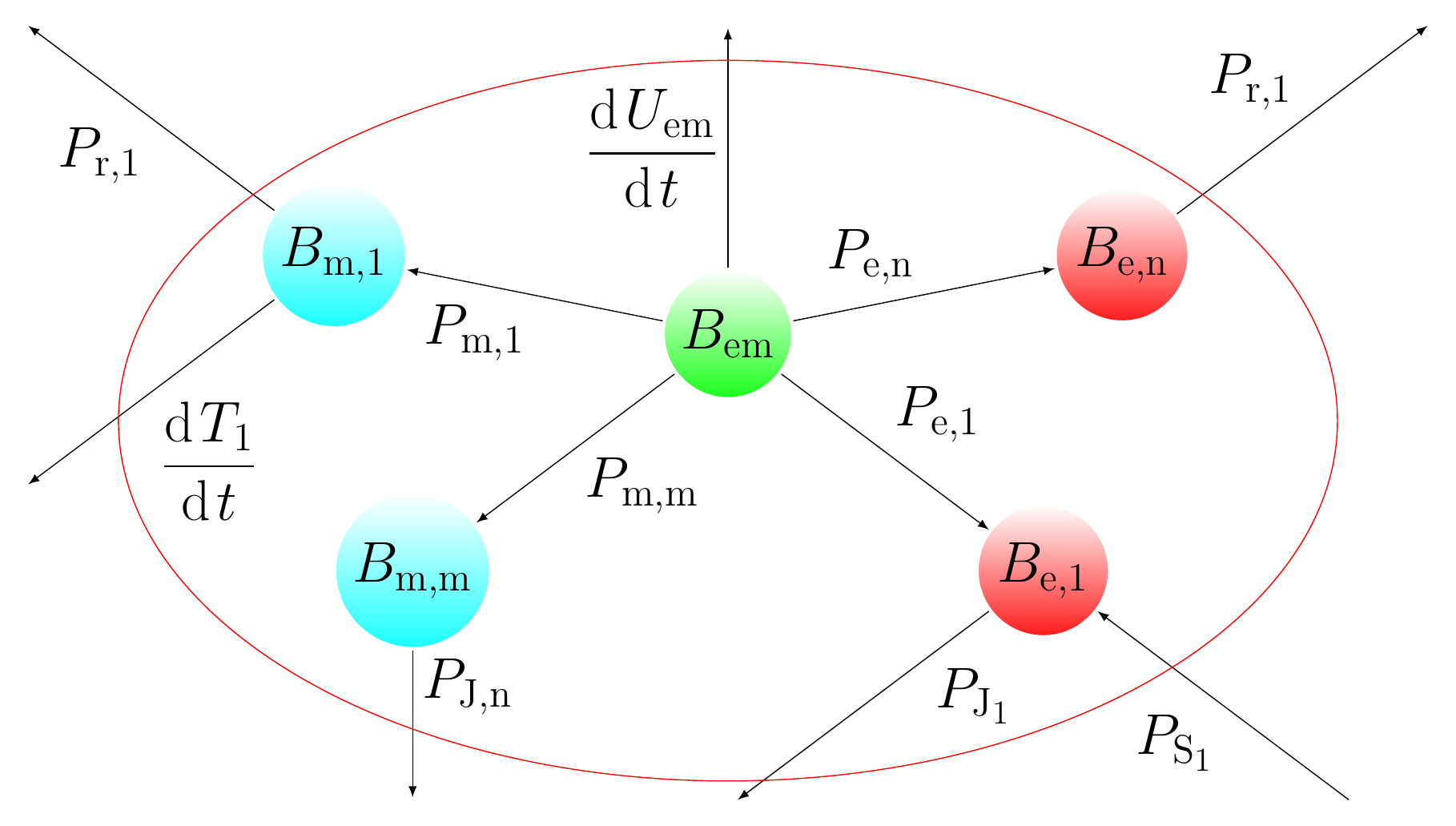}\end{center}
\caption{By selecting a group of nodes in the graph a new equation is obtained adding the powers represented by all the external edges. If all the nodes are selected, we get a global energy audit of the system. }
\label{fig::elip}
\end{figure}

Fig.\ref{fig::all} includes the analysis of all power coming from the electromagnetic field. New equations can be obtained by adding those of any number of nodes. Internal power flows are canceled out. Fig.\ref{fig::elip} represents the equation coming from all the nodes inside the closed red curve, which yields a {\em global} power audit of the system.

\begin{figure}[h]
\begin{center}\includegraphics[width=0.75\textwidth]{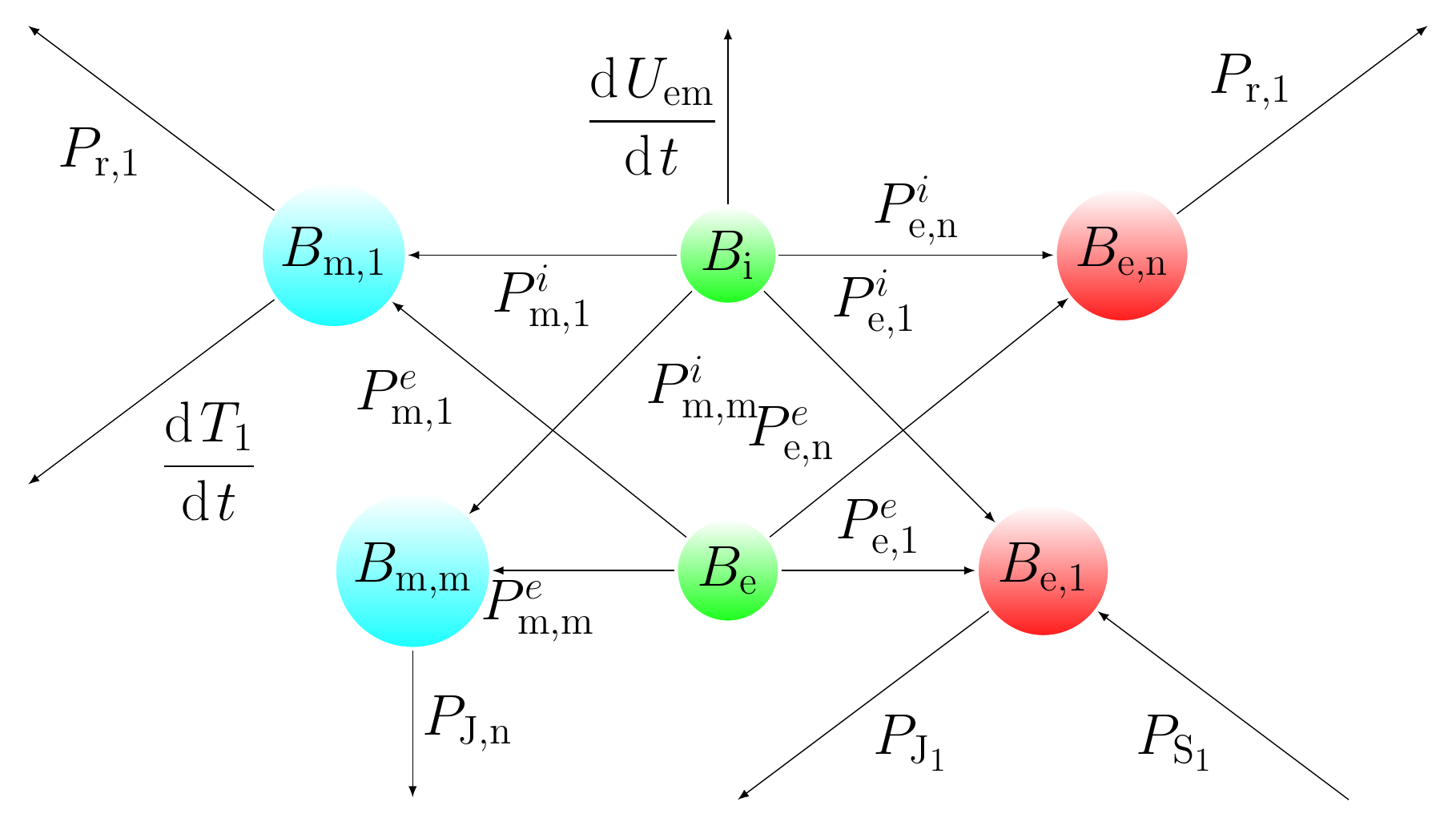}\end{center}
\caption{When a system is under constant magnetic fields generated elsewhere, two balances emerge. One is for the electrical and mechanical powers induced by the {\em local} magnetic fields and the other accounts for the cancellation of the electrical and mechanical powers combined coming from the external sources.}
\label{fig::dos}
\end{figure}

When there is an external source of constant magnetic fields, we have proved in Section \ref{sec::ext} that there is another 
balance equation , represented by Eq.\ref{qloc}. Besides, electric amd mechanical powers from the constant magnetic field cancel out. It is represented in Fig.\ref{fig::dos}.

\section{Examples}\label{sec::example}

\subsection{Variable reluctance circuit}
\begin{figure}[H]
\begin{center}\includegraphics[width=0.65\textwidth]{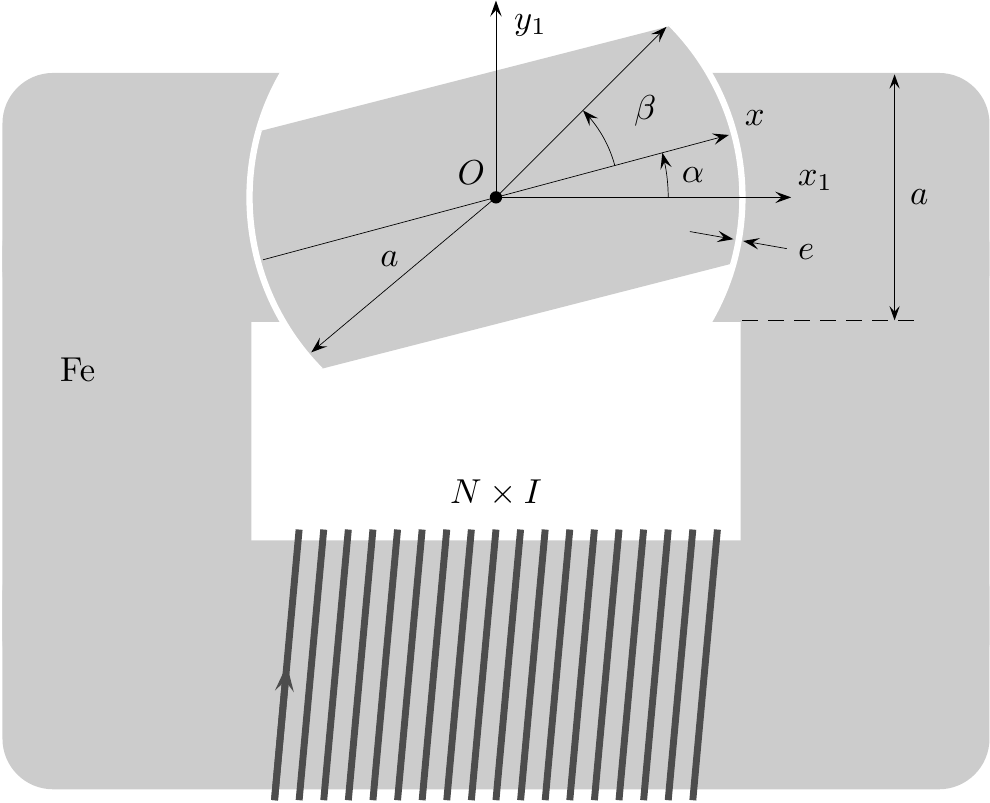}\end{center}
\caption{An electromechanical system consisting of a variable reluctance magnetic circuit. The magnetomotive force is supplied by a current $I$ wound $N$ times around the core. The rotating part makes the reluctance dependent on the rotation angle $\alpha$. The mechanical forces from the magnetic field tend to minimize the reluctance.}
\label{fig::rc}
\end{figure}

As a first example we may consider the system of Fig.\ref{fig::rc}. An electric source supplies an electric current $I$ which is wound $N$ times around the ferromagnetic core of a magnetic circuit. Part of the circuit may rotate by an angle $\alpha$, thus making the reluctance ${\cal R}(\alpha)$ variable.  It can be approximated by
\bel{relu}
{\cal R}(\alpha) = \frac {2 e}{\mu_0 ha(2\beta-\alpha)}
\eel
where $a,e,\beta$ are geometrical parameters read from Fig.\ref{fig::rc} and $h$ is the {\em height} (unseen dimension) of the system. 

The electromotive force, from Faraday's law results
\bel{emff}
{\cal E}=- N^2 \der{I {\cal R}^{-1}}{t} = -N^2 {\cal R}^{-1}\der{I}{t} - N^2 I\derp{{\cal R}^{-1}}{\alpha}\dot\alpha 
\eel
so that the electrical power supplied to the circuit by the field reads
\bel{epq}
P_{\rm e,1} = -N^2 {\cal R}^{-1} I \der{I}{t} - N^2I^2\derp{{\cal R}^{-1}}{\alpha}\dot\alpha 
\eel
The electromagnetic energy stored in the field is
\bel{Uoo}
U_{\rm em} =F= \frac{N^2I^2}{2 {\cal R}(\alpha)}
\eel
The mechanical torque results
\bel{mt}
F_\alpha = \derp{F}{\alpha}= \frac{N^2I^2}{2}\derp{ {\cal R}^{-1}(\alpha)}{\alpha}
\eel
giving a mechanical power
\bel{mp}
P_{\rm m,1}=F_{\alpha}\dot\alpha = \frac{N^2I^2}{2}\derp{ {\cal R}^{-1}(\alpha)}{\alpha}
\eel
It is now straightforward to check that Eqs.\ref{epq},\ref{mp} and the time derivative of Eq.\ref{Uoo} add to zero, which indicates that the energy balance matches.

\begin{figure}[H]
\begin{center}\includegraphics[width=0.65\textwidth]{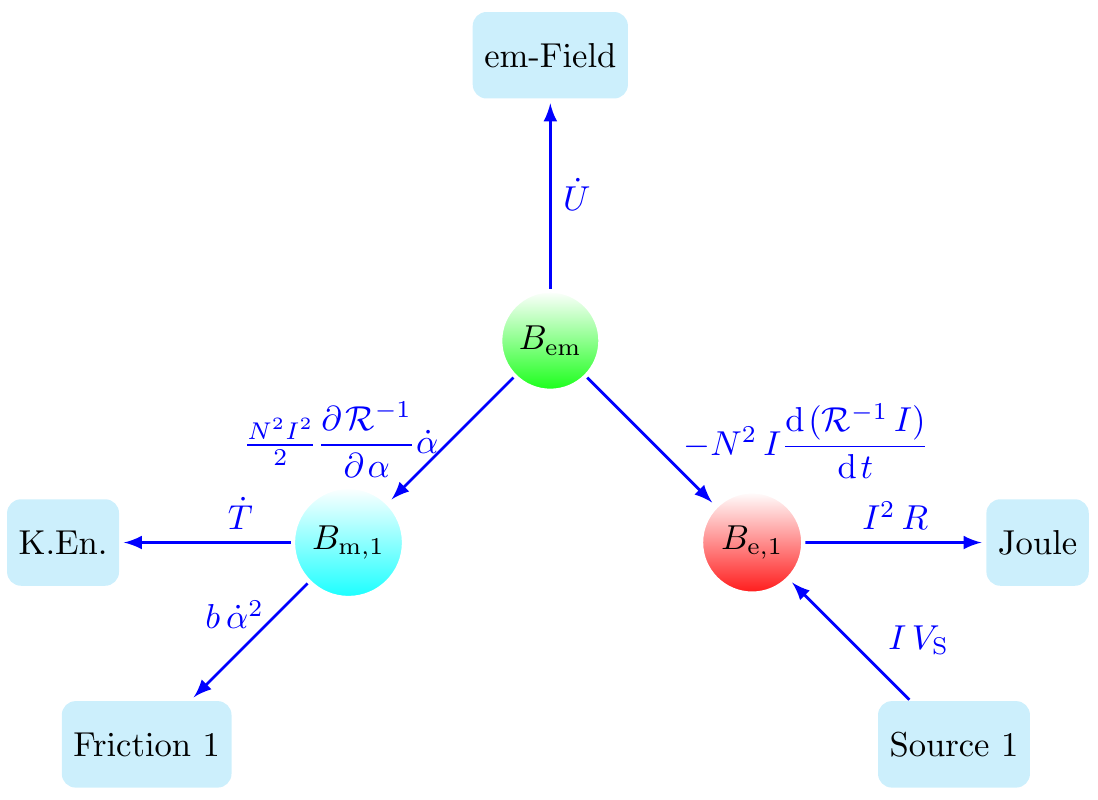}\end{center}
\caption{Power flowchart for the variable reluctance system of Fig.\ref{fig::rc}. Power from the source is traced down to the increase of kinetic energy, mechanical friction, Joule dissipation and energy storage in the field.}
\label{fig::rcpc}
\end{figure}

Then electrical and mechanical powers may enter further balances. Typically, there is a power furnished by the electrical source $V_s I$ and a Joule dissipation in the resistance of the wire $R I^2$. There may also be a gain in kinetic energy $\dot T$ of the rotary part and a mechanical viscous friction dissipation $b\dot \alpha^2$, as represented in the flowchart of Fig.\ref{fig::rcpc}. 

\begin{figure}[H]
\begin{center}\includegraphics[width=0.75\textwidth]{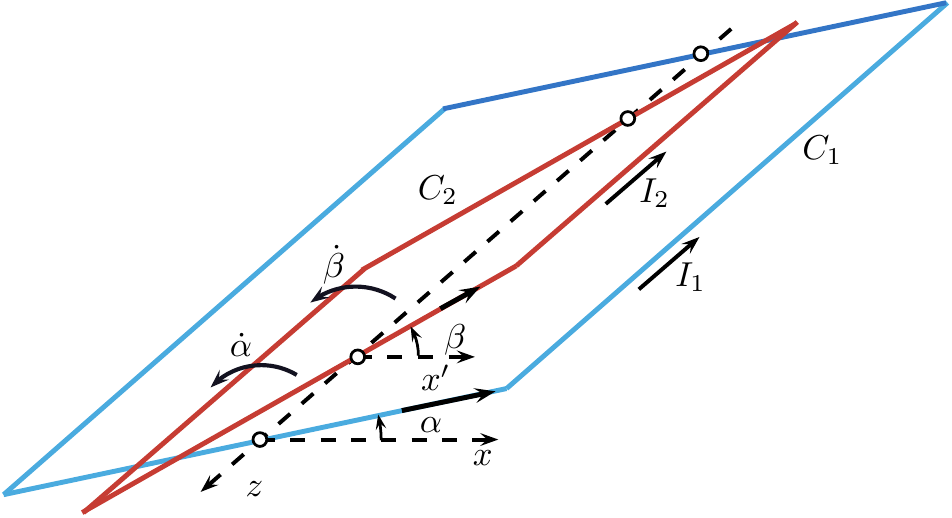}\end{center}
\caption{An electromechanical system of circuits $C_1,C_2$ which rotate about axis $z$ through angles $\alpha,\beta$. Their currents induce magnetic fluxes on each other. The mutual fluxes depend on currents and positions, hence the electromechanical interactions for which a power flowchart  is depicted in Fig.\ref{fig::bal}.}
\label{fig::two}
\end{figure}
\subsection{Electromagnetic clutch}
We may next consider the electromechanical system of Fig.\ref{fig::two} made by two electric circuits $C_1,C_2$ which may rotate about a fixed axis $z$, through angles $\alpha,\beta$ respectively. Both circuits may have voltage sources $V_{S1},V_{S2}$, resistances $R_1,R_2$ and are traversed by  electric currents $I_1,I_2$, respectively. The mechanical rotations encounter  viscous friction torques $ -b\dot\alpha,-b\dot \beta$, and may increase the kinetic energies $T_1,T_2$.  The mutual induction matrix is
\bel{matr}
\overline{\overline{M}} = \left(
\begin{array}{cc}
L &  M \cos \left( \beta-\alpha\right)\\M \cos \left( \beta-\alpha\right)& L
\end{array}
\right)
\eel

\begin{figure}[H]
\begin{center}\includegraphics[width=\textwidth]{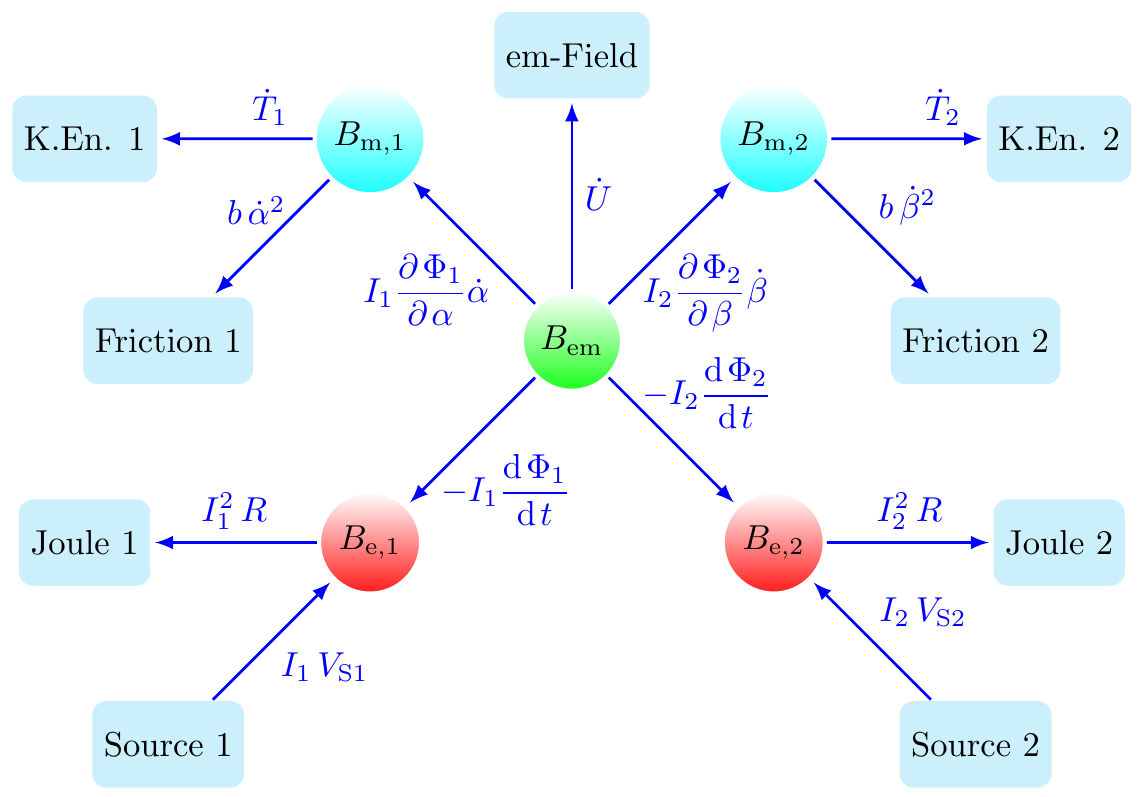}\end{center}
\caption{Power flowchart for the system of Fig.\ref{fig::two}. The power delivered (or absorbed) by the electric sources may be traced down to dissipative losses in resistors and viscous frictions and kinetic energy gains in the rotating masses.}
\label{fig::bal}
\end{figure}

Now we can evaluate the electromotive forces and electrical powers for both circuits, using Faraday's law
\bel{bothce}
\left(\begin{array}{c}
{\cal E}_1\\{\cal E}_2
\end{array}\right)=-\der{\,}{t}\left(\begin{array}{c}
L I_1 + MI_2 \cos(\beta-\alpha)\\L I_2 + MI_1 \cos(\beta-\alpha)
\end{array}\right) \eel
\bel{bothced}
\left(\begin{array}{c}
{\cal E}_1\\{\cal E}_2
\end{array}\right)
= \left(\begin{array}{c}
-L \dot I_1 + MI_2 \sin(\beta-\alpha)(\dot\beta-\dot \alpha) - M \dot I_2 \cos(\beta-\alpha)\\-L \dot I_2 + MI_1 \sin(\beta-\alpha)(\dot\beta-\dot \alpha) - M \dot I_1 \cos(\beta-\alpha)
\end{array}\right) 
\eel
\bel{bothp}
\left(\begin{array}{c}
P_{\rm e,1}\\P_{\rm e,2}
\end{array}\right)
= \left(\begin{array}{c}
-L I_1\dot I_1 + MI_1I_2 \sin(\beta-\alpha)(\dot\beta-\dot \alpha) - M I_1\dot I_2 \cos(\beta-\alpha)\\-L I_2\dot I_2 + MI_1I_2 \sin(\beta-\alpha)(\dot\beta-\dot \alpha) - M \dot I_1I_2 \cos(\beta-\alpha)
\end{array}\right) 
\eel
The mechanical torques and powers, according to Eq.\ref{ra}\footnote{In this problem energy and coenergy are the same, because the system is linear}, are 
\bel{bomQ}
\left(\begin{array}{c}
F_1\\F_2
\end{array}\right)
= \left(\begin{array}{c}
MI_1I_2 \sin(\beta-\alpha)\\-MI_1I_2 \sin(\beta-\alpha)
\end{array}\right) 
\eel
\bel{bomp}
\left(\begin{array}{c}
P_{\rm m,1}\\P_{\rm m,2}
\end{array}\right)
= \left(\begin{array}{c}
MI_1I_2 \sin(\beta-\alpha)\dot\alpha\\-MI_1I_2 \sin(\beta-\alpha)\dot\beta
\end{array}\right) 
\eel
The electromagnetic energy stored in the field is
\bel{Us}
U_{\rm em}=\frac{1}{2}\left(
L I_1^2 + 2 M I_1 I_2\cos(\beta-\alpha) + L I_2^2
\right)
\eel
and its time derivative evaluates to
\bel{Ust}
\der{U_{\rm em}}{t}=
L I_1 \dot I_1 - M I_1 I_2\sin(\beta-\alpha)(\dot\beta-\dot\alpha)+ M \dot I_1 I_2\cos(\beta-\alpha)+M I_1 \dot I_2\cos(\beta-\alpha) + L I_2 \dot I_2
\eel
It is now straightforward to  check Eq.\ref{bal} adding the powers of  Eqs.\ref{bothp},\ref{bomp} and \ref{Ust}  add to zero. The electrical and mechanical powers delivered by the field enter further power balances. Electrical powers may be added to those supplied by other sources and dissipate in resistors. Mechanical powers may increase kinetic energies and dissipate by frictional forces.  A complete power flowchart for the system is represented in Fig.\ref{fig::bal}. We can select all nodes to write down the equation:

\bel{glau}
I_1 V_{S1}+I_2V_{S2}=\dot U + \dot T_1+\dot T_2 + I_1^2 R_1 + I_2^2 R_2 + b\dot\alpha^2 + b \dot \beta^2
\eel

\printbibliography
\end{document}